# Influence of Strand Design, Boron Type, and Carbon Doping Method on the Transport Properties of Powder-in-Tube MgB$_{2-X}$C$_X$ Strands

Y. Yang[1], M. Susner[1], M. D. Sumption[1], M. Rindfleisch[2], M. Tomsic[2], E. W. Collings[1]

[1]Center for Superconducting and Magnetic Materials, Department of Materials Science and Engineering, The Ohio State University, Columbus, OH, USA

[2]Hypertech Research, Columbus OH, USA

**Abstract**

The transport properties of a number of MgB$_2$ strands have been investigated in terms of their response to strand design, starting B powder choice, and the approach to C doping used. The strands had various designs, specifically; (i) several chemical barriers were introduced, viz: Fe and Nb, (ii) the strands were encased in various outer-sheath materials, viz.: monel, Cu+monel, monel+glidcop, Nb+monel, (iii) the filament counts were varied (1, 18, and 36), and (iv) the final strand diameter was varied. In addition, for a subset of the strand designs several B powder and C-dopant types were investigated. Specifically, two types of amorphous B powder were used: (i) Moissan based "Tangshan boron" from the Tangshan Weihao Magnesium Powder Co. Ltd, China, (ii) "SMI-boron" from Specialty Metals Inc, USA, which is produced in a plasma torch by the reduction-by-hydrogen of BCl$_3$. Two approaches to C doping were taken: (i) "malic-acid treatment" in which C is introduced into the B powder precursor by the moderate temperature drying out a slurry of B mixed in with a malic-acid-toluene solution (during which the malic acid decomposes leaving C as the only solid residue) before the Mg powder is mixed in; (ii) direct C doping of the SMI-produced B by introducing a known percentage of CH$_4$ into the plasma flame. Critical current densities, $J_c$, were measured on 1.5 m long samples at 4.2 K in fields of up to 14 T; of all the strands measured, that doped with SMI-C at a nominal 4 mol% C (in relation B) yielded the highest $J_c$ values e.g $1.1\times10^5$ A/cm$^2$ at 7 T, $4.5\times10^4$ at 10 T, and $2.2\times10^4$ A/cm$^2$ at 12 T. The n-values are given for all strands at 5 and 10 T, and for a certain set of strands the magnetic field dependencies of the *n*-values and the influence of C-doping is presented. Finally we demonstrate that, over a wide range of *B*, log($J_c$) decreases linearly with *B* with a slope $-\alpha$ such that the $J_c(B)$ of any strand can be parameterized in terms of $\alpha$ and its zero-field intercept $J_c(B=0)$.

# 1. Introduction

MgB$_2$ superconducting materials and strands are relatively simple to make, are available at a reasonable cost, and have performance specifications that make them of interest for a number of applications. Notable among these is MRI, although other devices are also of interest including fault current limiters, motors, generators, and various special applications. There are three well known approaches to the fabrication of MgB$_2$ strand: (i) *ex-situ*, a powder-in-tube (PIT) process in which the tube, or sheath, is loaded with pre-reacted MgB$_2$ powder [1]; (ii) the "internal Mg diffusion" (IMD) process, an *in-situ* one in which the contents of the sheath consists of B powder surrounding a solid axial rod of Mg [2,3]; and (iii) the standard *in-situ* PIT process in which a chemically inert tube (barrier) filled with mixed B and Mg powders, is encased in an outer sheath, drawn to wire, and reacted for short times at moderate temperatures [4-6]. Numerous efforts at MgB$_2$ wire development have been undertaken [1-11], with good result.

In previous reports we have focused on the *in-situ* approach to MgB$_2$ strand fabrication, reporting on monofilamentary and multifilamentary strands, both in terms of their properties as materials [12], and as "conductors" [13,14]. We have reported on the development of *in-situ* multifilamentary strands with up to 54 filaments whose basic design embodies Nb barriered filaments in a Cu matrix, all encased in an outer Monel or Cu-Ni sheath. In the present paper we describe their transport properties ($J_c$ and *n*-value) in response to variation of strand design details and starting powder optimization. Specifically investigated are; (i) the outer-sheath material, (ii) the interfilamentary matrix material (iii) the strand final diameter, (iv) the number of filaments, (v) the starting B powder, and (vi) the choice of C-bearing dopant.

We have reported previously on property differences stemming from the choice of B [15], as have other groups [16, 17]. In general, for *in-situ* conductors, the smaller the B particle size, the better the transport $J_c$. In this work we investigate two of the higher performing B types, one Moissan based, the other made via a plasma spray technique. It is also known that C and C-bearing dopants (including SiC) are some of the best dopants for enhancing the lower temperature performance of $MgB_2$. Various methods have been used to introduce C, including SiC [18,19], direct C additions [20], various hydrocarbons [21,22], and numerous organic compounds [23,24]. They all have similar effects although small differences in efficiency can be found, generally small scale and very uniform additions are most effective. The "limit" to the C-doping process is the addition of C to the starting B powder as in the SMI process described below. This leads to very high properties in the final strand as described in [25]. Strands formed from powder similar to that of [25] have been fabricated in longer lengths; their properties are described in further detail in this work in terms of $J_c$ and *n*-value.

Below we first describe the strands under investigation. After that we characterize their transport properties at 4.2 K over a range of applied fields. The results in terms of transport $J_c$ and *n*-value are then discussed with reference to strand design, B-power type, and C-doping method.

## 2. Experimental

### 2.1. Strand Fabrication

A series of 30 monofilamentary and multifilamentary strands (the latter with 18 and 36 filaments) were fabricated by HyperTech Research, Inc (HTR) using the by-now well known CTFF process (a variant of the Powder in Tube process [5]). Most of the strands were 0.83 mm in diameter; they included a thick chemical barrier (typically of Nb but in some cases Fe), and outer sheaths which were either Monel 400® (henceforth "monel", a nickel-copper alloy) or monel associated with Cu, GlidCop® (henceforth "glidcop", an oxide-dispersion-strengthened Cu), or Nb. Their specifications, structures, and heat treatments are listed in Table 1. The basic powder ingredients were commercial Mg powders (99%, ≈ 20-25 μm particle size) and B powder from one of two sources: (i) "Tangshan boron" from the Tangshan Weihao Magnesium Powder Co. Ltd, China, produced using the Moissan process (ii) "SMI-boron" from Specialty Metals Inc, USA, produced in a plasma torch by the reduction-by-hydrogen of $BCl_3$. Also included from time to time in the starting B powders were small percentages of the dopant C, introduced: (i) by the moderate temperature drying out of a slurry of B mixed in with a malic-acid-toluene solution, during which the malic acid decomposes leaving C as the only solid residue [23,26], and (ii) direct C doping of the SMI-produced B by including a known percentage of $CH_4$ into the plasma flame [27]. After being drawn to size the strands were heat treated at temperatures of $675^oC$ to $750^oC$ for times of 20 to 120 min as seen in Table 1.

## 2.2. Strand Design

Niobium was used as the "chemical barrier" material for most of the wires described here since it has minimal reaction with the Mg and B powders during the reaction heat treatment. Iron was used as a chemical barrier (as a replacement for Nb) for a few strand types, but required some intermediate annealing during wire drawing. Monel was the outer sheath for most strands because of its high flow strength in combination with its ductility, although it was in a few cases laminated with glidcop or pure Cu in the interests of electrical stability. Microstructure images were obtained for several samples using a Sirion field emission SEM in backscatter mode. Figure 1 (a) shows strand Type A, a simple monofilament with Nb barriers and an outer monel sheath. Figures 1 (b)-(g) show the various multifilamentary strand geometries, which are described in Table 1. Strand Types B and C (Table 1, Figure 1) are 18 and 36 filament variants of Nb-chemical barrier strands with the Nb-clad filaments packed together inside a monel sheath. In this case no Cu matrix separates the Nb filaments as in previous strand designs [13,28]. Instead a central Cu filament (Nb for design style B*) was placed in the center of the multifilamentary arrangement, for both mechanical and electrical reasons -- (i) to minimize centerburst (an instability related to flow stress gradients through the strand diameter during wiredrawing), and (ii) to aid strand stability. Strand Type D consisted of 36 filaments in a double wall sheath of Cu and monel (and a central Cu filament), while strand Type E, with 18 superconducting filaments plus a central Cu filament, used a monel/glidcop double wall sheath. Strand Type F which consisted of 18 filaments protected by Fe chemical barriers, had no Cu matrix, and used a monel outer sheath. Strand Type G had a Nb wrap around the Fe-clad filaments but otherwise was similar to strand Type F.

## 2.3. Measurements

Transport $J_c$ measurements were performed on all samples at 4.2 K in pool boiling liquid helium in transverse magnetic fields ranging from 0 T to 15 T. Measurements were made not on "short-samples", but rather on 1.5 m long samples helically wound on modified "ITER barrel" holders (i.e. 32 mm diameter Ti-Al-V formers [29]). The gauge length was 500 mm, and the electric field criterion for transport $J_c$ was 1 µV/cm.

## 3. Results for Critical Current Density

The results of $J_c$ measurements on MgB$_2$ strands of various designs and powder compositions are given in three sections. Section 3.1 is primarily a **B-source comparison**. It compares the $J_c$s of strands produced from SMI-boron with and without directly doped 1%C (and also an additional 2% malic acid) with those from Tangshan-boron treated with 5% malic acid. Section 3.2 deals with the effect on $J_c$ of changes in **strand architecture** (barrier and sheath compositions, filament number, strand diameter) and reaction heat treatment of strands all based on SMI-boron that had been directly doped with a nominal 2%C -- i.e. "SMI-C2%". In Section 3.3, which also deals with strands based on SMI-C-doped boron we report on the variation of $J_c$ with **change of C content** in the starting B. Tables are included that compare the "low-field" (i.e. 5 T) and "high-field" (i.e. 10 T) 4.2-K $J_c$s of all the strands.

### 3.1. Malic-Acid-Doped Strands based on two Types of B Powder

The two B-powder types used in the fabrication of malic-acid-doped strands were Tangshan B (TsB) and plasma spray powders (SMI). Two plasma spray powders were used, one with no C added during the plasma spray process (A-SMI-Malic2%)-1F), and one with 1% C added during plasma spray (B*-SMI-C1%-Malic2%). In both cases 2% malic acid was added, see Table 1. Here the 2%malic refers to a nominal mol% of malic in the final MgB$_2$ compound, and the C1% related to the SMI powder strands refers to a nominal molar addition of C to the B in the gas stream. However, the actual molar % of C relative to the final MgB$_2$ compound is listed for all strands in column 3 of Table 1. Carbon levels for all distinct mixtures of SMI C-doped powder and malic acid doped powder have been confirmed experimentally with a C-analysis performed by LECO. We note that this C analysis only measures the total amount of C present in the sample, and does not assess the level to which the C has entered the B sublattice. The two SMI strands had in the finish 0.8mol%C (A-SMI-Malic2%-1F), and 1.7mol% (B*-SMI-C1%-Malic2%). Also included in this comparison were two strands with 4.77 mol% malic acid doped TsB boron (leading to a final C content of 1.6 mol% in the $MgB_2$), an 18 stack Type B with Nb barriers and an 18 stack Type G with Fe barriers. The $J_c$ versus $B$ curves for these samples are shown in Figure 2, and Table 2 lists the strands of Figure 2 in descending order of 5 T (left three columns) and 10 T (right three columns) 4.2 K $J_c$. The SMI B samples with 2% malic doping in addition to 1% C pre-doping (B*-SMI-C1%-Malic2%) have greater high field $J_c$ values than does the sample with malic acid doping and non-pre-doped SMI B (A-SMI-malic2%-1F), due to greater C doping level (2x). At lower fields the C-doping is less important, although the sample with the higher temperature HT performs better. Specifically, at 4.2K the $J_c$ of B*SMI-C1%-Malic-750/30 is about $10^5 A/cm^2$ at 5 T and more than $10^4 A/cm^2$ at 10 T. This may suggest that malic acid doped SMI samples prefer higher heat treatment temperatures. The Tanshan B-based samples, although they have a similar level of C to the B*-SMI-C1%-malic2% samples, perform similarly to the lower C doped sample A-SMI-malic 2%-1F. This might be either because of smaller grain sizes which are known to be present for the SMI powders, or because of a difference in the amount of C uptake into the lattice [25]. For the TsB-based samples, the best transport $J_c$ was again obtained for the higher reaction temperature. Heat treated for 30min/700°C the TsB-malic strand attained a $J_c$ of $5x10^4 A/cm^2$ at 5 T and more than $10^4 A/cm^2$ at 8T. We also note that the Fe barrier sample showed lower $J_c$ value especially at high fields. We note that in this set of six samples there is no obvious relationship between $J_c$ and $n$. Finally, to a first approximation, critical current density has an exponential field dependence and hence can be expressed in the form

$$J_c = J_{c0} \cdot \exp(-B/B_0) \quad (1)$$

In which $J_{c0}$ is the zero-field $J_c$ and $B_0$ is a fitting parameter (not directly related to the upper critical field). It follows that

$$LnJ_c = LnJ_{c0} - \alpha B \quad (2)$$

in which $\alpha = 1/B_0$. In this model $LnJ_c$ decreases linearly with $B$ with slope $-\alpha$; hence the $J_c$ field dependence of a class of superconductors can be characterized in terms of a single parameter, $\alpha$. If, as is typical, $J_c$ was plotted on a based 10 log vs linear plot, the slope $\gamma = 0.434\alpha = 0.434/B_0$. The $B_0$ and $J_{c0}$ values extracted for the first set of data are shown in Table 3. This fit works well below a field $B^*$, also listed in Table 3.

### 3.2. Strand Design

Figure 3 (a) shows transport $J_c$ field dependence at 4.2 K in response to variation of strand design, viz: various chemical barrier and sheath materials, filament numbers, and strand diameters. In order to emphasize "design", all strands used the same powder mixture, specifically SMI B with a nominal 2% C addition (see Table 1 and Ref [25]) and the same Mg-B ratio of 1:2 (with the exception of strand Type F which is B-rich, see Table 1). On the other hand, a range of HTs is included. Figure 3 (b) is an expanded region of Figure 3 (a), and Table 4 lists the 5 T and 10 T $J_c$s and $n$-values in order of decreasing $J_c$. First, it can be noted that most of the strands have 5 T $J_c$s greater than $10^5$ A/cm$^2$, the highest $J_c$ in the set being 1.78 x $10^5$ A/cm$^2$. The monofilament, Type A, had a lower performance than most of the multifilament strands, with the exception of the Fe-barrier multifilament strand. This Fe-barrier strand, Type F, which had a B-rich powder composition, had the lowest 5 T $J_c$, 6.6 x $10^4$ A/cm$^2$. Type B strands are represented throughout Table 4. On the other hand Type C strands are clustered near the top of the table (higher $J_c$ performance), while Types E and D are near the middle (with A and F closer to the bottom, as noted above). Type B (18 filaments) and C (36 filament) strands are both monel sheathed with Nb barriers. Strand Types D and E have composite sheaths: that of Type D being Cu-monel, and Type E being monel-glidcop. It may be that the extra strength of the pure monel sheath as compared to the composite sheaths allows for better pre-compression during cooldown (from the reaction temperature), as was seen to be important for Bi-based superconducting strands previously [30]. The fact the that monofilamentary Type A underperforms all of the multifilamentary types, suggests that the additional core-compaction associated with multifilamentary re-stacking is important, and/or that the larger metal-to-powder ratio of multifilamentary strands, leading to greater compression during cool–down[30] is also important. On this subject we also note that reducing the diameter of B*-SMI-C2%-700/120 from 0.984 mm to 0.834 mm increased 10T $J_c$ from 1.61 up to 2.08 (x $10^4$ A/cm$^2$), and that reducing the diameter of E-SMI-C2%-MG from 1.008 mm to 0.834 mm increased 10T $J_c$ from 1.4 to 1.6 (x $10^4$ A/cm$^2$).

Strand Type B experienced the four HTs 675°C/20 minutes, 675°C/60 minutes, 700°C/60 minutes, and 700°C for 120 minutes, Table 4. No discernable relationship between HT and either the 5 T or 10 T $J_c$s can be seen. In fact, recent studies completed in this laboratory have shown the optimum HT temperatures to be about 675°C-700°C (high enough to insure complete reaction, low enough to minimize grain growth and unwanted reactions). The properties are relatively insensitive to HT duration once the full reaction is reached (about 20 minutes at these temperatures). After removing the variables of strand design, barrier type, powder type, filament count, and strand diameter, Type B appears to have a $J_c$ variation of about 30% (defined as [$J_{c,min}$-$J_{c,ave}$]/$J_{c,ave}$). Strand Type C is represented by only one HT in Table 4, namely 700°C/60 minutes. If we do a similar estimation of its variation, we obtain 12.5%.

Returning to Figures 3 (a) and (b) we note that the $J_c(B)$ slopes are similar, not unexpected since the powder mixtures and Mg/B ratios are the same (except for Strand F). If we assume a $J_c$ field dependence of the form $J = J_{c0}\exp(-B/B_0)$, $B_0$ ranges from 4.5 to 6.0 T. Finally, in Table 4 we find no particular relationship between $J_c$ and $n$-value; to be expected for a group of variously designed strands.

### *3.3. Critical Current Density of C-Doped Plasma Spray-Boron in Response to Variation of Carbon Content*

The $J_c$s of strands fabricated from SMI-boron doped with three different levels of carbon are depicted in Figure 4 and presented in order of decreasing 5 T and 10 T $J_c$ in Table 5. As reported in [25] (for short 3 cm samples) the strands designated SMI-C1%, SMI-C2%, and SMI-C4% had measured C levels of 1.29 mol%, 2.10 mol%, and 4.0 mol%, respectively. These strands were monofilamentary, 0.83 mm OD, with a Nb chemical barrier and an outer monel sheath (further strand details are available in ref [25]). Figure 4 shows $J_c$ increases rapidly with increasing levels of C doping. The two SMI-C4%-based strands have $J_c$s of more than $10^4$ A/cm$^2$ at 13 T – higher than that of HTR's best SiC-doped strands. Irreversibility field measurements on these strands [25] indicate that SMI-C4% has about the optimal level of C doping. The $J_c(B)$ results depicted in Figure 4 clearly indicate their division into three groups depending on carbon content. At 10 T the $J_c$ of the SMI-C4% pair, at $4\times10^4$ A/cm$^2$, is an order of magnitude higher than that of the SMI-C1% pair and that of SMI-C2% has an intermediate value. An increase of the HT temperature from 675°C to 700°C (both for 60 min) uniformly raises the $J_c$ of SMI-C1% but produces no changes in the $J_c$s of the SMI-C2% and SMI-C4% strands. The $B_0$, $J_{c0}$, and $B^*$ values extracted for this set of data are shown in Table 6, where a clear increase in $B_0$ is seen with C-additions.

## 4. *n*-values of MgB$_2$ Strands`

While many studies of critical current density for MgB$_2$ have been performed, the *n*-value (or index number) is less frequently reported. Nevertheless the results of some studies with *n*-value as a central focus have been reported. Measuring *in-situ* HTR-fabricated strands Flukiger et al [31] found *n*-values of about 5 at 8 T, 4.2 K, rising to 20-30 at 4 T, 4.2 K. They noted an unspecified but non-linear variation of *n* with *B*, and that densification of the strands substantially improved the *n*-values. Goldacker et al [32], measuring both *in-situ* and *ex-situ* wires, saw an exponential field dependence, with *n*-values of 10-20 at 8 T, 4.2 K, and 20-40 at 4T, 4.2 K. Kitaguchi et al [33], measuring *in-situ* processed strands, obtained *n*-values of 17 at 10 T, 4.2 K, values which increased to 27 with SiC doping additions. Suo [34] et al, achieved *n*-values of 15-30 at 8 T, 4.2K, and above 60 at 4T, 4.2 K. Martinez [35] using magnetization measurements to extract *n*-values, found apparently empirical correlations of *n*-value to $J_c$. Similar correlation of *n*-value and $J_c$ were seen in the work of Kim et al [36], among samples where values of about 30 were seen at 8T, 4.2K. It should be noted that all of these studies were performed on short samples, and for that reason we might expect that they would be less susceptible to extrinsic limitations of *n* value. As a limit, the intrinsic *n*-value is determined by the pinning potential [37-39]. In any case, shorter samples should be less susceptible to the extrinsic *n*-variations brought on by $I_c$ variations of larger wavelength (large compared to the sample size).

As stated above the present transport $J_c$ measurements were performed on 1.5 m long samples helically wound on modified "ITER barrels"; *n*-values were obtained from the accompanying voltage-current data within the electric field range of 0.4 μV/cm to 4 μV/cm. The 5 T and 10 T results for all the strands are presented in Tables 2, 4, and 5. The $J_c$ results for the strands with various SMI-C doped starting powders shown in Figure 4 are complemented by the n-values shown in Figure 5. Here we note several different behaviors for *n* as a function of field. For some samples the *n* vs *B* curve is relatively flat, while for others n increases with field. It seems that samples with higher C content may have a better *n*-value, although the trend is not uniform. It appears that some mixture of intrinsic and extrinsic contribution to *n*-values may be present.

In order to make more general use of this data, it is helpful useful to remember that the basic index number relationship is

$$E = E_c \left( \frac{J}{J_c} \right)^n \tag{3}$$

This allows the curvature of the *I-V* curve to be described, where $E_c$ is a given electric field criterion, $J_c$ is the *J* for that criterion, and *E* and *J* are the electric field and current density. If we chose to use a different criterion, say

$$E = E_c' \left( \frac{J}{J_c'} \right)^n \tag{4}$$

Then

$$\frac{J_c'}{J_c} = \left( \frac{E_c'}{E_c} \right)^{1/n} \tag{5}$$

Using this expression we predict the $J_c$ value associated with a new electric field criterion given a $J_c$ at one criterion and the associated index number.

Consider now two strands with different *n*-values but which are otherwise identical. Let one of the strands have an infinite *n*-value. This strand will transition to the normal state with infinite sharpness, at a current density we can define as $J_\infty$. If we let the second strand have some finite *n*-value, then the $J_c$ of this strand using the same electric field criterion, $E_c$, will be lower than the first. Let us consider the condition of this strand at $J = J_\infty$. Here the electric field will be that of the wire at the transition to the normal state, which we can take to be $E_m$. Eq (1) then becomes

$$\frac{E_m}{E_c} = \left(\frac{J_\infty}{J_c}\right)^n \tag{6}$$

which can be re-written

$$J_c = J_\infty \frac{1}{\left(\frac{E_m}{E_c}\right)^{1/n}} \tag{7}$$

or alternatively

$$Ln J_c = Ln J_\infty - \frac{1}{n} Ln\left(\frac{E_m}{E_c}\right) \tag{8}$$

i.e

$$Ln J_c = A - \frac{1}{n} D \tag{9}$$

Where A and D are constants. These last two equations express the fact that for a given strand, given all other factors (specifically including the intrinsic $J_c$, $J_\infty$) being equal, as $n$ decreases so does $J_c$.

## 5. Discussion and Conclusions

The transport properties of a number of $MgB_2$ strands have been investigated in terms of their response to strand design, starting B powder choice, and the approach to C doping used. In general, it was seen that multifilamentary strands, and strands with stronger outer sheaths had a higher $J_c$ performance. This may be due to the tendency of the outer sheath to apply a pre-stress during cool-down after HT thereby densifying the superconducting core of the strand and improving connectivity. Wires with smaller diameters tended to perform better, at least within a limited range of diameters. The improvement with strand diameter reduction may again be the result of core densification. Little variation with HT was seen, in the small window defined by 675-700°C and 20-120 minutes, a temperature-time range already seen to be optimum for previously measured strands of similar design.

In addition, B powder and C-dopant types were investigated, namely a fine Moisson type B (Tangshan boron) and a plasma spray based B (SMI-B). Generally plasma spray B performed better than even small powder type Moisson B. C doping was added both directly to the B (in the Plasma Spray process) or after the fact using malic acid additions. Both powder types responded well to malic acid treatment. However, the best C doping performance was seen after direct C doping (i.e., C doped into the starting SMI-process B), and the optimum amount for 4 K operation is presently set at 4 mol%.

Critical current densities, $J_c$, were measured on 1.5 m long samples at 4.2 K in fields of up to 14 T; of all the strands measured, that doped with SMI-C at a nominal 4 mol% C yielded the highest $J_c$ values e.g $1.1 \times 10^5$ A/cm$^2$ at 7 T, $4.5 \times 10^4$ at 10 T, and $2.2 \times 10^4$ A/cm$^2$ at 12 T. The n-values were given for various strands and reached 20 at 4 T. In addition, we parameterized the strands in terms of slopes $\alpha$ and zero-field intercept $J_c(B=0)$ values.

**Tables**

Table 1.  Strand Specifications.

Table 2. 4.2 K transport $J_c$ values of SMI-boron strands doped with a nominal 2% malic acid and Tangshan-boron strands doped with a nominal 5% malic acid at 5T and 10T, respectively.

Table 3. $B_0$, $B^*$, and $J_{c0}$ values of SMI-boron strands doped with a nominal 2% malic acid and Tangshan-boron strands doped with a nominal 5% malic acid.

Table 4. 4.2 K transport $J_c$ values of SMI-boron strands directly doped with a nominal 2% C with various strand designs and heat treatments at at 5T and 10T, respectively.

Table 5. 4.2 K transport $J_c$ values of SMI-boron strands directly doped with nominal levels of 1%C, 2%C, and 4%C at 5T and 10T, respectively.

Table 6. $B_0$, $B^*$, and $J_{c0}$ values of SMI-boron strands directly doped with nominal levels of 1%C, 2%C, and 4%C.

Table 1

| Name | HTR Tracer No | Actual C (mol %)[c] | Chemical Barrier | Sheath | Fil.Count | HT(C°/min) | %SC | OD (mm) |
|---|---|---|---|---|---|---|---|---|
| *SMI Boron Samples* | | | | | | | | |
| A-SMI-C2%-1F-675/20 | 2035 | 2.3% | Nb | M | 1 | 675/20 | 15.8 | 0.834 |
| A-SMI-Malic2%-1F-700/20[a] | 1980 | 0.8% | Nb | M | 1 | 700/20 | 16 | 0.834 |
| | | | | | | | | |
| B*-SMI-C1%-Malic2%-675/30[a] | 2061 | 1.7% | Nb | M | 18 | 675/30 | 21.3 | 0.834 |
| B*-SMI-C1%-Malic2%-750/30[a] | 2061 | 1.7% | Nb | M | 18 | 750/30 | 21.3 | 0.834 |
| B*-SMI-C2%-675/20 | 2066 | 2.3% | Nb | M | 18 | 675/20 | 17.4 | 0.834 |
| B*-SMI-C2%-700/120-a | 2066 | 2.3% | Nb | M | 18 | 700/120 | 17.4 | 0.834 |
| B*-SMI-C2%-700/120-b | 2066 | 2.3% | Nb | M | 18 | 700/120 | 19 | 0.984 |
| B*-SMI-C2%-700/120-c | 2097 | 2.3% | Nb | M | 18 | 700/120 | 17.4 | 0.834 |
| | | | | | | | | |
| B-SMI-C1%-675/20 | 2110 | 1.4% | Nb | M | 18 | 675/20 | 23.6 | 0.834 |
| B-SMI-C1%-700/20 | 2110 | 1.4% | Nb | M | 18 | 700/20 | 23.6 | 0.834 |
| B-SMI-C2%-675/60-a | 2115 | 2.3% | Nb | M | 18 | 675/60 | 22.7 | 0.834 |
| B-SMI-C2%-675/60-b | 2163 | 2.3% | Nb | M | 18 | 675/60 | 25.8 | 0.834 |
| B-SMI-C2%-700/120 | 2115 | 2.3% | Nb | M | 18 | 700/120 | 20.1 | 0.984 |
| B-SMI-C2%-700/60 | 2163 | 2.3% | Nb | M | 18 | 700/60 | 25.8 | 0.834 |
| B-SMI-C2%-T-675/60 | 2115T | 2.3% | Nb | M | 18 | 675/60 | 22.7 | 0.83 |
| B-SMI-C2%-T-700/60 | 2115TA | 2.3% | Nb | M | 18 | 700/60 | 22.7 | 0.83 |
| B-SMI-C4%-675/60 | 2158 | 4.8% | Nb | M | 18 | 675/60 | 15.4 | 0.834 |
| B-SMI-C4%-700/60 | 2158 | 4.8% | Nb | M | 18 | 700/60 | 15.4 | 0.834 |
| | | | | | | | | |
| C-SMI-C2%-36F-700/60-a | 2154R | 2.3% | Nb | M | 36 | 700/60 | 15.5 | 0.83 |
| C-SMI-C2%-36F-700/60-b | 2148 | 2.3% | Nb | M | 36 | 700/60 | 17.3 | 0.83 |
| C-SMI-C2%-36F-700/60-c | 2148P1 | 2.3% | Nb | M | 36 | 700/60 | 17.8 | 0.94 |
| C-SMI-C2%-36F-700/60-d | 2148P2 | 2.3% | Nb | M | 36 | 700/60 | 16.7 | 0.93 |
| | | | | | | | | |
| D-SMI-C2%-CuM-36F-675/60 | 2189 | 2.3% | Nb | Cu/M | 36 | 675/60 | 16.1 | 0.934 |
| D-SMI-C2%-CuM-36F-700/60 | 2189 | 2.3% | Nb | Cu/M | 36 | 700/60 | 16.1 | 0.934 |
| | | | | | | | | |
| E-SMI-C2%-MG-700/60-a | 2170B | 2.3% | Nb | M/G | 18 | 700/60 | 11.4 | 0.834 |
| E-SMI-C2%-MG-700/60-b | 2170D | 2.3% | Nb | M/G | 18 | 700/60 | 10.2 | 1.008 |
| | | | | | | | | |
| F-SMI-C2%-Fe-Brich-675/20 | 2020 | 2.3% | Fe | M | 18 | 675/20 | 15.4 | 0.83 |
| | | | | | | | | |
| *Tangshan Boron (TsB)+Malic-Acid-Treated Samples* | | | | | | | | |
| B-TsB-Malic5%-675/60[b] | 2056 | 1.6% | Nb | M | 18 | 675/60 | 14.3 | 0.83 |
| B-TsB-Malic5%-700/30[b] | 2056 | 1.6% | Nb | M | 18 | 700/30 | 14.3 | 0.83 |
| | | | | | | | | |
| G-TsB-Malic5%-Fe-NbM-675/60[b] | 2017 | 1.6% | Fe | Nb/M | 18 | 675/60 | 13.2 | 1.008 |

*Explanation of the Sample Name:*
(1) The prefixes "A, B*, B, C – G" refer to the various strand architectures depicted in Figure 1
(2) The letters "a", "b", "c" and "d" attached to otherwise "similar" samples designate variations in %SC and strand OD as listed. The letter 'T' indicates twisted.
(3) A Nb chemical barrier is understood, otherwise an iron barrier is indicated by "Fe".
(4) A monel sheath is understood, otherwise monel+glidcop, copper+monel. niobium+monel are indicated by M/G, Cu/M and Nb/M, respectively.
(5) A filament count of 18 is understood, otherwise counts of 1 and 36 are indicated by 1F and 36F, respectively.
(6) "$B_{rich}$" indicates extra boron hence $MgB_{2.5}$.

a Here malic2% refers to mol% in the final MgB2 compound, but does not presuppose complete substitution of the C into the B sublattice.
b Here malic5% refers to mol% in the final MgB2 compound, but does not presuppose complete substitution of the C into the B sublattice.
c Here mol % is relative to the final MgB2 compound, but does not presuppose complete substitution of the C into the B sublattice.

Table 2.

| $J_c$ at 5T($10^4$ A/cm$^2$) | $n$-value | Name | $J_c$ at 10T($10^4$ A/cm$^2$) | $n$-value | Name |
|---|---|---|---|---|---|
| 9.6 | 6.7 | B*-SMI-Malic2%-750/30 | 1.6 | 10.9 | B*-SMI-Malic2%-750/30 |
| 6 | 5.9 | A-SMI-Malic2%-1F-700/20 | 1 | 12.9 | B*-SMI-Malic2%-675/30 |
| 5.5 | 20.5 | B*-SMI-Malic2%-675 /30 | 0.61 | 3.5 | B-TsB-Malic5%-700/30 |
| 5 | 8.1 | B-TsB-Malic5%-700/30 | 0.53 | 2.4 | B-TsB-Malic5%-675/60 |
| 4.3 | 11.2 | G-TsB-Malic5%-Fe-NbM-675/60 | 0.5 | 1.9 | A-SMI-Malic2%-1F-700/20 |
| 4.2 | 5.1 | B-TsB-Malic5%-675/60 | 0.23 | 6.5 | G-TsB-Malic5%-Fe-NbM-675/60 |

Table 3.

| Name | $B_0(T)$ | $J_{c0}(10^6 A/cm^2)$ | $B*(T)$ |
|---|---|---|---|
| A-SMI-Malic2%-1F-700/20 | 1.9 | 1.1 | 6 |
| B*-SMI-Malic2%-675/30 | 2.1 | 1.2 | 8 |
| B*-SMI-Malic2%-750/30 | 2.0 | 2.4 | 8 |
| B-TsB-Malic5%-675/60 | 2.3 | 0.4 | 5 |
| B-TsB-Malic5%-700/30 | 1.6 | 1.1 | 3 |
| G-TsB-Malic5%-Fe-NbM-675/60 | 1.6 | 1.1 | 5 |

Table 4.

| $J_c$ at 5T($10^4$ A/cm$^2$) | $n$-value | Name | $J_c$ at 10T($10^4$ A/cm$^2$) | $n$-value | Name |
|---|---|---|---|---|---|
| 17.8 | 12.0 | B*-SMI-C2%-700/120-a | 2.08 | 14.8 | B*-SMI-C2%-700/120-a |
| 17.5 | 6.5 | B*-SMI-C2%-675/20 | 2.06 | 10.2 | C-SMI-C2%-36F-700/60-b |
| 17.4 | 7.6 | C-SMI-C2%-36F-700/60-c | 2.04 | 8.9 | B-SMI-C2%-675/60-b |
| 16.8 | 14.5 | B-SMI-C2%-675/60-b | 1.99 | 9.1 | C-SMI-C2%-36F-700/60-c |
| 16.5† | -- | B-SMI-C2%-700/60 | 1.98 | 9.1 | C-SMI-C2%-36F-700/60-d |
| 16.3 | 8.6 | C-SMI-C2%-36F-700/60-a | 1.97 | 10.6 | B*-SMI-C2%-675/20 |
| 16 | 6.2 | C-SMI-C2%-36F-700/60-d | 1.87 | 7.9 | C-SMI-C2%-36F-700/60-a |
| 15.3 | 8.3 | C-SMI-C2%-36F-700/60-b | 1.84 | 8.4 | B-SMI-C2%-700/60 |
| 14.2 | 9.2 | E-SMI-C2%-MG-700/60-a | 1.61 | 3.3 | E-SMI-C2%-MG-700/60-a |
| 13.3 | 9.6 | E-SMI-C2%-MG-700/60-b | 1.6 | 5.1 | A-SMI-C2%-1F-675/20 |
| 12.5 | 12.2 | D-SMI-C2%-CuM-36F-675/60 | 1.47 | 7.3 | D-SMI-C2%-CuM-36F-675/60 |
| 12 | 11.2 | B-SMI-C2%-T-675/60 | 1.43 | 4.9 | B*-SMI-C2%-700/120-c |
| 11.1 | 14.7 | B-SMI-C2%-700/120 | 1.42 | 12.7 | B-SMI-C2%-T-675/60 |
| 10.2 | 10.5 | D-SMI-C2%-CuM-36F-700/60 | 1.4 | 4.6 | E-SMI-C2%-MG-700/60-b |
| 9.8 | 14.9 | B-SMI-C2%-675/60-a | 1.27 | 7.6 | B-SMI-C2%-700/120 |
| 9.5 | 12.9 | A-SMI-C2%-1F-675/20 | 1.27 | 6.3 | B*-SMI-C2%-700/120-b |
| 9.4 | 6.9 | B*-SMI-C2%-700/120-c | 1.26 | 3.9 | B-SMI-C2%-T-700/60 |
| 9.2 | 7.4 | B-SMI-C2%-T-700/60 | 1.2 | 6.5 | D-SMI-C2%-CuM-36F-700/60 |
| 8.2 | 3.4 | B*-SMI-C2%-700/120-b | 1.04 | 5.8 | B-SMI-C2%-675/60-a |
| 6.6 | 8.7 | F-SMI-C2%-Fe-Brich-675/20 | | | |

Note:" † " indicates data is extrapolated from fitting curve of the other data , not measured data.

Table 5.

| $J_c$ at 5T($10^4$ A/cm$^2$) | $n$-value | Name | $J_c$ at 10T($10^4$ A/cm$^2$) | $n$-value | Name |
|---|---|---|---|---|---|
| 20.0† | -- | B-SMI-C4%-700/60 | 4.6 | 10.7 | B-SMI-C4%-700/60 |
| 19.9 | 21.2 | B-SMI-C4%-675/60 | 4.1 | 14.5 | B-SMI-C4%-675/60 |
| 17.1 | 14.5 | B-SMI-C2%-675/60-b | 2.1 | 8.9 | B-SMI-C2%-675/60-b |
| 16.0† | -- | B-SMI-C2%-700/60 | 1.9 | 8.4 | B-SMI-C2%-700/60 |
| 6.4 | 4.3 | B-SMI-C1%-700/20 | 0.7 | 5.7 | B-SMI-C1%-700/20 |
| 4.6 | 9.5 | B-SMI-C1%-675/20 | 0.4 | 6.9 | B-SMI-C1%-675/20 |

Note:" †" indicates data is extrapolated from fitting curve of the other data , not measured data.

Table 6.

| Name | $B_0(T)$ | $J_{c0}(10^6 A/cm^2)$ | $B^*(T)$ |
|---|---|---|---|
| B-SMI-C1%-675/20 | 1.6 | 1.6 | 7 |
| B-SMI-C1%-700/20 | 1.7 | 2.0 | 8 |
| B-SMI-C2%-675/60-b | 1.8 | 5.0 | 8 |
| B-SMI-C2%-700/60 | 1.8 | 4.7 | 8 |
| B-SMI-C4%-675/60 | 2.8 | 1.5 | 8 |
| B-SMI-C4%-700/60 | 2.7 | 1.8 | 9 |

**Figure Captions**

Figure 1. SEM-backscatter images of a set of representative strands. (a) *Strand Type A:* Single Nb filament (monocore) in a Nb barrier sheathed in monel; (b) *Strand Type B:* 18 Nb-clad (barrier) filaments (leading to 18 $MgB_2$ filaments imbedded in Nb), plus a central solid-Cu filament, all enclosed in a monel outer sheath -- the strands prefixed B* in Table 1 have a central solid-Nb filament instead, see (b*); (c) *Strand Type C:* 36 Nb-clad filaments (leading to 36 $MgB_2$ filaments imbedded in Nb), plus a central Cu filament, all enclosed in a monel outer sheath; (d) *Strand Type D:* 36 Nb-clad filaments, plus a central Cu filament, all enclosed in a Cu-inner/monel-outer double-wall sheath designated Cu/M in Table 1; (e) *Strand Type E:* 18 Nb-clad filaments, plus a central Cu filament, all enclosed in a monel-inner/glidcop-outer double-wall sheath designated M/G in Table 1; (f) *Strand Type F:* 18 Fe-clad filaments (leading to 18 $MgB_2$ filaments imbedded in Fe), plus a central "CTFF-formed" Fe filament, enclosed in a monel outer sheath; (g) *Strand Type G:* 18 Fe-clad filaments (leading to 18 $MgB_2$ filaments imbedded in Fe), plus a central "CTFF-formed" Fe filament, enclosed in a Nb-inner/monel-outer double-wall sheath designated Nb/M in Table 1.

Figure 2. 4.2 K transport $J_C$ versus $B$ for SMI-boron strands doped with a nominal 2% malic acid and Tangshan-boron strands doped with a nominal 5% malic acid.

Figure 3. (a) 4.2 K transport $J_c$ versus $B$ for SMI-boron strands directly doped with a nominal 2% C with various strand designs and heat treatments; (b) Detail for the field range 5-10 T.

Figure 4. 4.2 K transport $J_c$ versus $B$ for SMI-boron strands directly doped with nominal levels of 1%C, 2%C, and 4%C.

Figure 5. *n*-value versus $B$ for SMI-boron strands directly doped with nominal levels of 1%C, 2%C, and 4%C.

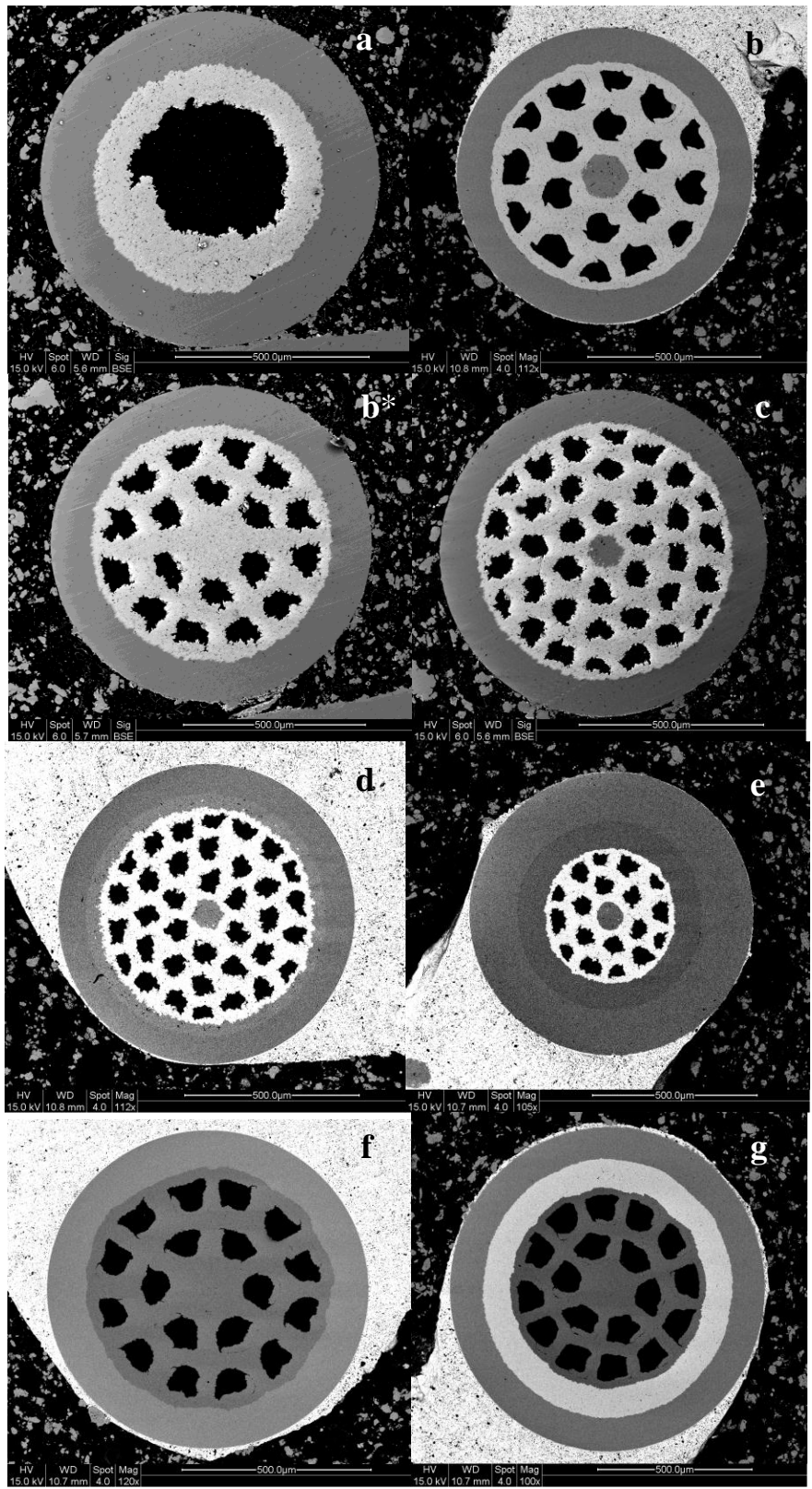

Figure 1

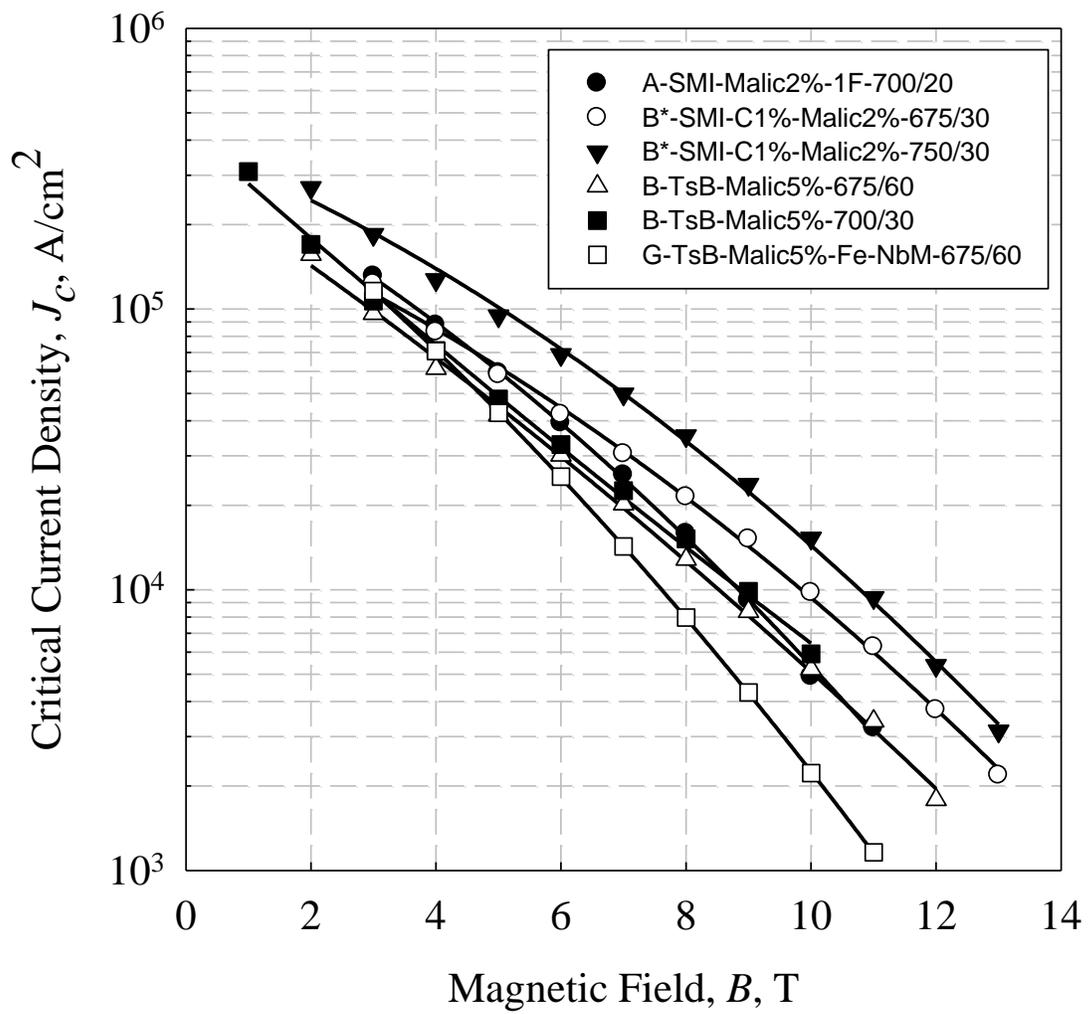

Figure 2

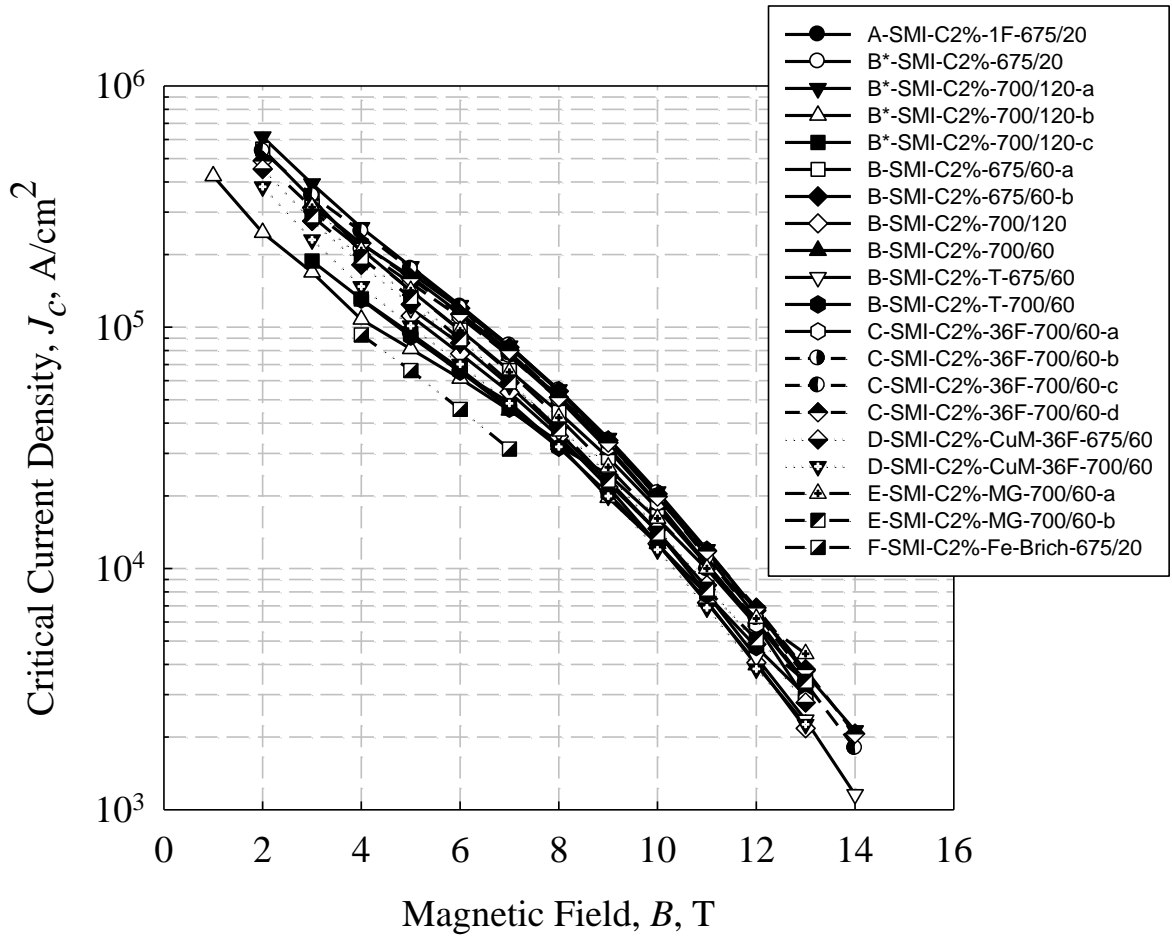

Figure 3(a)

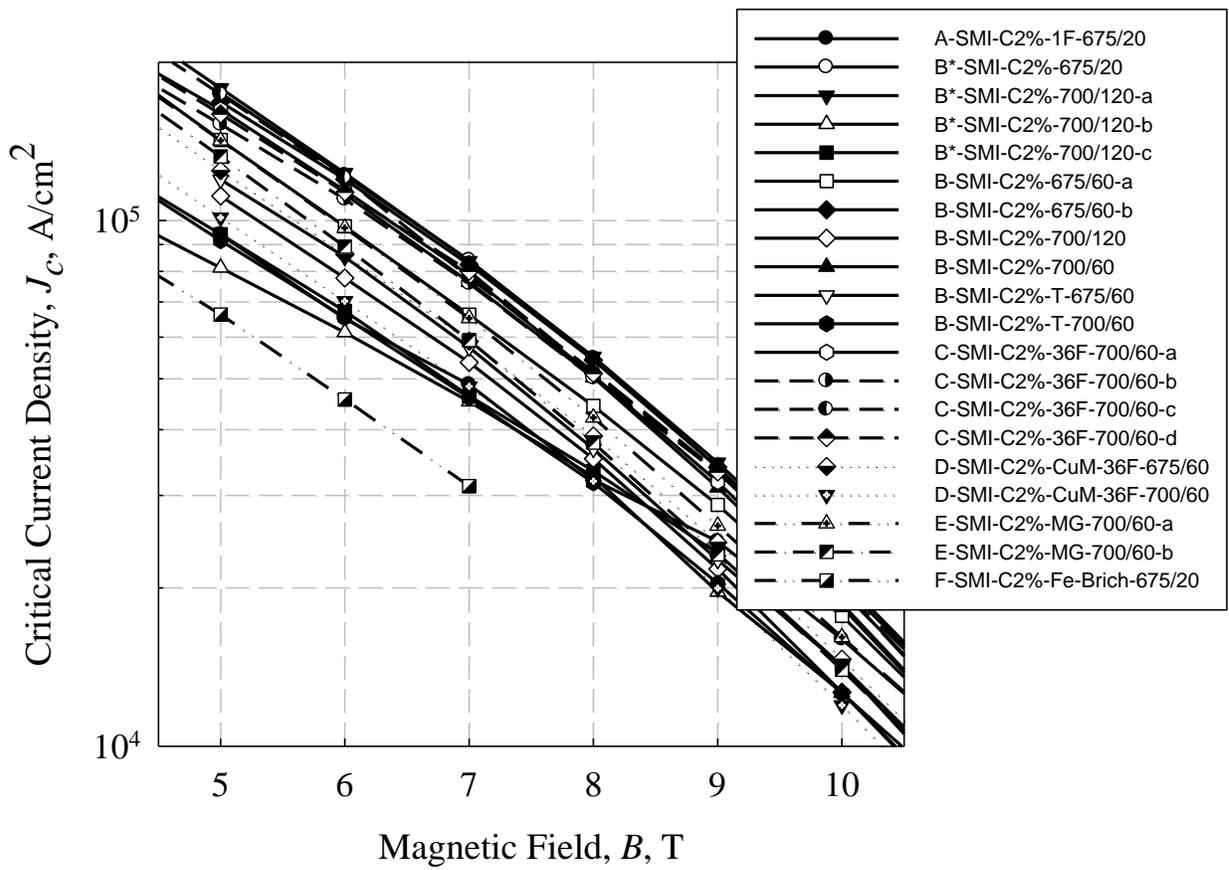

Figure 3(b)

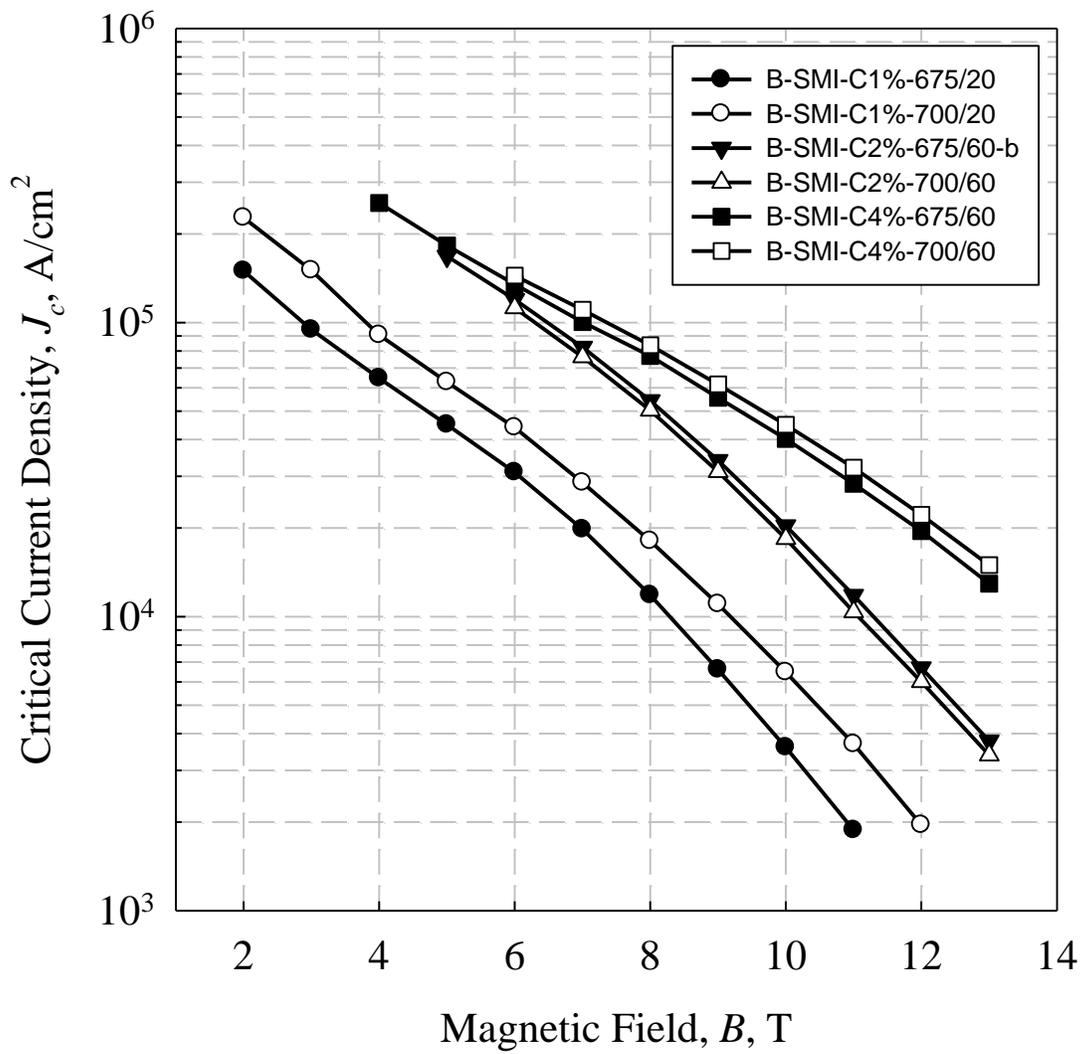

Figure 4

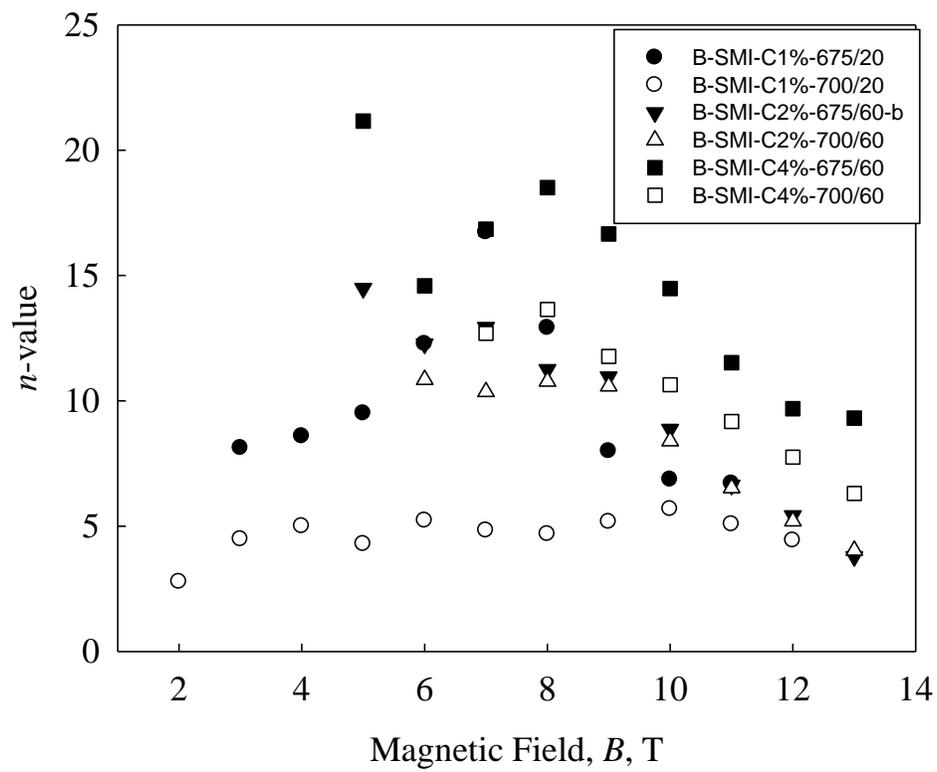

Figure 5